# Vortex Dynamics and Entropic Forces in Ising and Potts Antiferromagnets and Ice Models

Cristopher Moore[*], Mats G. Nordahl[†], Nelson Minar[‡], and Cosma Shalizi[*,§]

[*]*Santa Fe Institute, 1399 Hyde Park Road, Santa Fe, NM 87501* {moore,shalizi}@santafe.edu
[†]*Institute of Theoretical Physics, Chalmers University of Technology, S-412 96 Göteborg, Sweden* tfemm@fy.chalmers.se
[‡]*MIT Media Lab, 20 Ames Street, Cambridge, MA 02139* nelson@media.mit.edu
[§]*Physics Department, University of Wisconsin, Madison, WI 53706*

We study the dynamics of topological defects in the triangular Ising antiferromagnet, a related model on the square lattice equivalent to the six-vertex ice model, and the three-state antiferromagnetic Potts model on the square lattice. Since each of these models has a height representation in which defects are screw dislocations, we expect them to be attracted or repelled with an entropy-driven Coulomb force. In each case we show explicitly how this force is felt through local fields. We measure the force numerically, both by quenching the system to zero temperature and by measuring the motion of vortex pairs. For the three-state Potts model, we calculate both the force and the defect mobility, and find reasonable agreement with theory.

## I. INTRODUCTION

Topological defects play a major role in mediating phase transitions. In this paper, we look at several models with discrete degrees of freedom: the triangular Ising antiferromagnet, a related model on the square lattice equivalent to the six-vertex ice model, and the $q = 3$ Potts antiferromagnet on the square lattice. In each model we review how to describe defects as vortices or screw dislocations in a height representation, and write the total charge inside a finite region as a sum around its perimeter.

In each case, we find that the motion of the defects is proportional to a local field $\vec{F}$. If the states far away from a defect are uncorrelated, then this field decreases as $B/2\pi r$ where $B$ is the Burgers vector of the defect. We therefore argue that defects are governed by a first-order Coulomb force,

$$\langle \Delta r \rangle \propto \vec{F} \propto \frac{qq'}{r}$$

acting like vortices in a viscous regime or dislocations in a solid.

To confirm this, we do two kinds of numerical experiments. In a quench to zero temperature, we measure the density of defects as a function of time, and find logarithmic corrections to $\rho \propto 1/t$ like those seen in the $XY$ model. However, a free diffusion experiment suggests that this occurs whenever a local conservation law is used to determine the defects' initial positions and density fluctuations, whether or not there are forces between them. Therefore, we also attempt to measure the force directly by placing a pair of defects $r$ apart, allowing the lattice around them to equilibrate, and measuring how they would move toward or away from each other if we allowed them to. We obtain good agreement with a force of the form $\langle \Delta r \rangle = A/(r + r_0)$, although finite-size effects and a partial lack of ergodicity at $T = 0$ affect our results.

Since the energy of a pair of defects does not depend on the distance between them, this force must be entropically driven. We review how both the height model and heuristic arguments give a term in the entropy proportional to $-\ln r$, whose gradient then gives a $1/r$ force. Using Park and Widom's calculation of the stiffness of the equivalent height model, we predict that $A$ for the Potts antiferromagnet should be $3/4$. This is in reasonable agreement with our results.

## II. THE TRIANGULAR ISING ANTIFERROMAGNET

The Ising antiferromagnet

$$U = \sum_{nn} s_i s_j$$

on the triangular lattice can be experimentally realized in anhydrous alums with a yavapaiite layered structure [8] such as RbFe(SO$_4$)$_2$. Its equilibrium behavior was solved by Wannier [35] and Houtappel [16]. Its critical point is at zero temperature, where it has a nonzero entropy per site of

$$\frac{2}{\pi} \int_0^{\pi/3} \ln(2 \cos \omega) \, d\omega = \frac{1}{\pi} \text{Im} \sum_{k=1}^{\infty} \frac{e^{\pi i k/3}}{k^2} \approx 0.323066$$

Stephenson [32] showed that its correlations at $T = 0$ decay algebraically as $r^{-1/2}$.

Since every triangular plaquet of the lattice must contain at least one frustrated bond, we define a defect as a plaquet where all three bonds are frustrated, i.e. where all three sites are in the same state. An isolated defect cannot be annealed away with local changes; as figure 1 shows, defects diffuse under single spin-flip dynamics, maintaining their orientation on the lattice while reversing their spins. Thus there are two types of defects, $\bigtriangledown$ and $\triangle$. We assign these charges of $+1$ and $-1$ respectively, and pairs of opposite type can annihilate when they meet. These defects seem to have been first noticed by Landau [23].

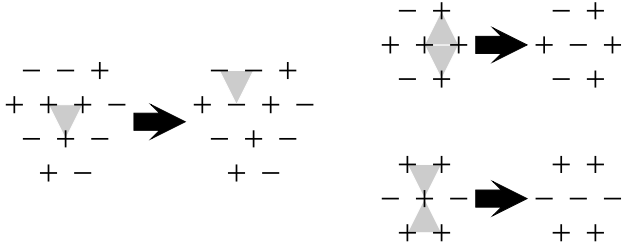

FIG. 1. Defects in the triangular antiferromagnet diffusing and annihilating.

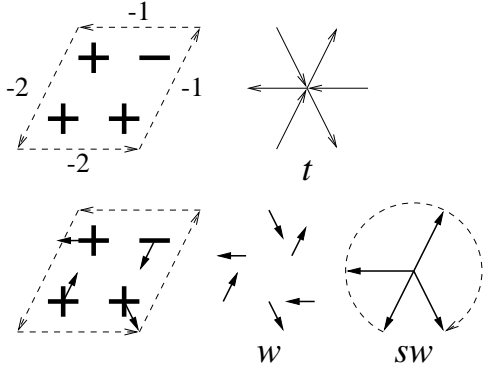

FIG. 2. Two ways of calculating the defect charge inside a region. Equation 1 gives $(1/6)(-2-2-1-1) = -1$, and $s_i w_i$ winds once clockwise around the origin. Both show a single negative ($\triangle$) defect.

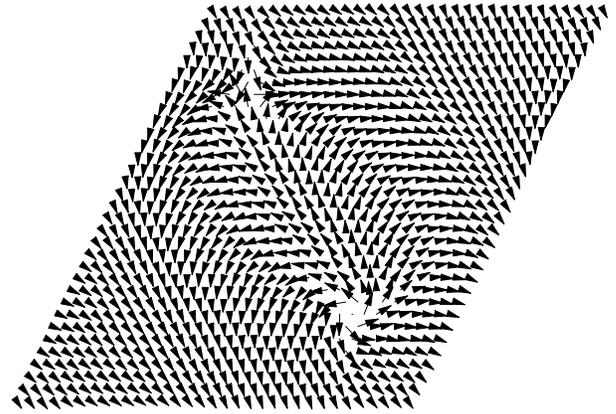

FIG. 3. The local magnetization $sw$ around a pair of defects. This picture was obtained by relaxing a $33 \times 33$ rhombic lattice until only two defects remained, and then holding them fixed for $10^5$ updates per site while we averaged the magnetization at each site. Finally, $sw$ was averaged over triplets of neighboring sites to smooth it.

We can define the charge within a region as an integral going counterclockwise along the bonds $\vec{b}$ of its perimeter,

$$Q = \frac{1}{6} \sum_{\vec{b}} \left( \frac{3}{2} s_i s_{i+1} + \frac{1}{2} \right) \vec{b} \cdot \vec{t} \qquad (1)$$

An example is shown in figure 2. Here $\vec{t}$ is a unit vector that lends an orientation to each bond in the lattice, and the expression in the parentheses is 2 for bonds connecting like sites and $-1$ for unlike ones. This latter quantity was used by Blöte et al. [7,26] as a height difference between neighboring sites to map the triangular antiferromagnet onto a solid-on-solid model of the $(1, 1, 1)$ corner of a simple cubic crystal, and this was generalized by Zeng and Henley [38] to arbitrary spin. Defects then correspond to screw dislocations with Burgers vector $\pm 6$.

Another way to define the charge is through a local magnetization, whose winding is like a vorticity [15]. We can three-color the sites of the lattice with vectors $2\pi/3$ apart, $w = e^{2\pi i k/3}$ for $k = 0, 1, 2$, such that moving positively along $\vec{t}$ rotates $w$ counterclockwise by $2\pi/3$. Then $Q$ is the winding number of $s_i w_i$ around the origin, which we can write as a contour integral if we like:

$$Q = \frac{1}{2\pi i} \oint \frac{d(sw)}{sw} \qquad (2)$$

This method is also shown in figure 2. In figures 3 and 4 we show the time average of $sw$ on a lattice with a single pair of defects. The fields around both defects, with opposite windings, are clearly visible.

If we define a field $\vec{E}$ by rotating $\vec{t}$ clockwise by $\pi/2$ and multiplying the result by the same quantity $(3/2)s_i s_{i+1} + (1/2)$ as in equation 1, then $\vec{E}$ will tend to point towards negative defects and away from positive ones. We can then rewrite equation 1 as

$$Q = \frac{1}{6} \sum_{\vec{b}} \vec{E} \times \vec{b} \qquad (3)$$

We can now establish a linear relationship between $\vec{E}$ and the expected motion of a defect. As figure 5 shows, once we choose which site of a positive defect to flip, the center of the new defect will be equal to the location of that site plus $\vec{E}/(2\sqrt{3})$ (we take the lattice spacing to be 1) where we measure $\vec{E}$ on the dashed edge facing the site being flipped. Since all three sites are equally likely, the average movement is

$$\langle \Delta \vec{x} \rangle = \frac{1}{6\sqrt{3}} \sum \vec{E} \qquad (4)$$

where the sum is over the three dashed edges surrounding the defect.

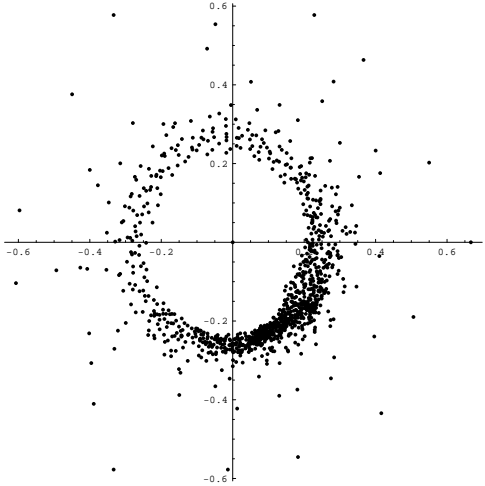

FIG. 4. A scatterplot of the averaged values of $sw$ shown in figure 3.

Now if we draw a perimeter of radius $r$ around a defect of charge $q$, and if we assume the system is isotropic, equation 3 implies that the average field is

$$\vec{E} = \frac{6q}{2\pi r}$$

pointing away from the defect. Here we assume that other defects are sufficiently far away so that they don't bias the surrounding lattice. If this is true, then combining it with equation 4 gives

$$\langle \Delta \vec{x} \rangle = \sqrt{3} \frac{qq'}{2\pi r}$$

for the expected motion of another defect of charge $q'$. In other words, we expect the average motion to be governed by a Coulomb force, in a viscous regime with constant mobility. The coefficient $\sqrt{3}/2\pi$ is only approximate, since we have used a local mean-field approximation in which $\vec{E}$ has no correlations between the dashed edges and so $(1/3) \sum \vec{E} = \bar{E}$.

Obviously, this will only hold if the field is sufficiently uncorrelated at this distance, and if the presence of another defect adds linearly to $\vec{E}$ rather than nonlinearly. For instance, if curvature of one defect's field is correlated with the presence of another, the field lines will be bunched between the two defects, and the force will fall off more slowly than $1/r$. In the sections below, we perform various numerical experiments to measure $\langle \Delta \vec{x} \rangle$, and we claim that it does in fact decrease as $1/r$.

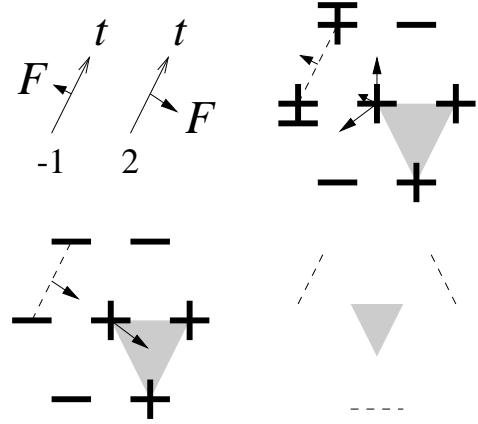

FIG. 5. Defining $\vec{E}$ and establishing its relationship with $\langle \Delta \vec{x} \rangle$. There are two cases where $\vec{E} = 1$ away from the defect, and averaging them gives $\langle \Delta \vec{x} \rangle = 1/(2\sqrt{3})$ towards the dashed edge. When $\vec{E} = 2$ towards the defect, the site does not flip since this would increase the energy by creating three defects instead of one, and relative to that site the defect moves $1/\sqrt{3}$ back to its original position. In the case not shown, both spins on the dashed edge are $+$, and the defect annihilates with a neighboring one. We can ignore this case for defects far apart.

## III. A RELATED MODEL ON THE SQUARE LATTICE: THE NO-STRIPES RULE AND THE SIX-VERTEX ICE MODEL

Consider the following model on the square lattice:

$$U = \sum_{\substack{s_1 \square s_2 \\ s_3 \square s_4}} -s_1 s_4 - s_2 s_3 + s_1 s_2 s_3 s_4$$

If we like, we can think of this as two Ising ferromagnets, one on each sublattice, strongly coupled by the four-point interaction like a kind of Ashkin-Teller model [1]. The Hamiltonian has the effect of giving plaquets of the form $\begin{smallmatrix} + & + \\ + & + \end{smallmatrix}$, $\begin{smallmatrix} + & - \\ + & - \end{smallmatrix}$, $\begin{smallmatrix} + & - \\ - & + \end{smallmatrix}$ and their rotations and reversals energies of $-1$, while giving plaquets of the form $\begin{smallmatrix} + & - \\ + & - \end{smallmatrix}$ and their rotations an energy of $+3$. We call these last plaquets 'stripes,' and they act as defects in the system. As figure 6 shows, isolated stripes diffuse and annihilate in pairs.

This model is related to the triangular antiferromagnet through the charge-preserving zig-zag transformation shown in figure 7. This preserves energy for the most part, with the one discrepancy that a defect-free square plaquet can be mapped onto an adjacent pair of positive and negative defects in the triangular lattice. Since such a pair will quickly annihilate, this should not significantly affect the dynamics.

Nevertheless, this 'no stripes' rule is quite elegant, and is worth studying on its own. There are two types of stripes, one which is vertical $\begin{smallmatrix} \pm & \mp \\ \pm & \mp \end{smallmatrix}$ when on even lattice squares and horizontal $\begin{smallmatrix} \pm & \pm \\ \mp & \mp \end{smallmatrix}$ when on odd ones, and the other which is vice versa. If we assign charges $+1$ and $-1$ to these respectively, the total charge $Q$ inside a region can be written as a sum over bonds $\vec{b}$ going counterclockwise around it,

$$Q = \frac{1}{4} \sum_{\vec{b}} \left( s_i s_{i+1} \times \left\{ \begin{array}{l} +1 \text{ if } \vec{b} \text{ is vertical} \\ -1 \text{ if } \vec{b} \text{ is horizontal} \end{array} \right\} \right.$$
$$\left. \times \left\{ \begin{array}{l} +1 \text{ if } \vec{b} \text{ has an even square on its left} \\ -1 \text{ if } \vec{b} \text{ has an odd square on its left} \end{array} \right\} \right) \quad (5)$$

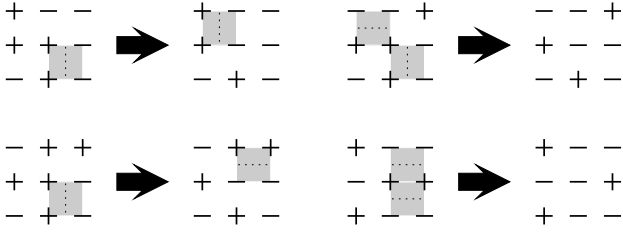

FIG. 6. Defects in the triangular Ising antiferromagnet diffusing and annihilating.

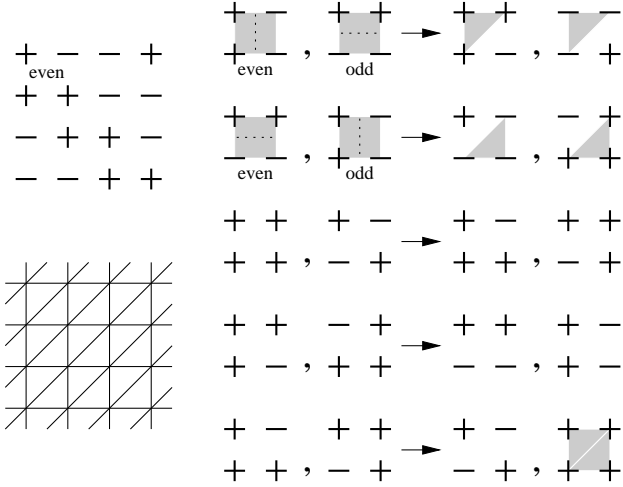

FIG. 7. By flipping half the sites in a zig-zag pattern, we can transform the 'no stripes' model into the triangular antiferromagnet in a charge-preserving way.

An example is given in figure 8.

If we define a field $\vec{E}$ perpendicular to each bond by rotating $\vec{b}$ clockwise or counterclockwise $\pi/2$ according to whether the quantity summed in equation 5 is positive or negative, then $\vec{E}$ is the same regardless of the orientation of $\vec{b}$. Moreover, $\vec{E}$ will tend to point toward negative stripes and away from positive ones, and we can rewrite equation 5 as

$$Q = \frac{1}{4} \sum_{\vec{b}} \vec{E} \times \vec{b}$$

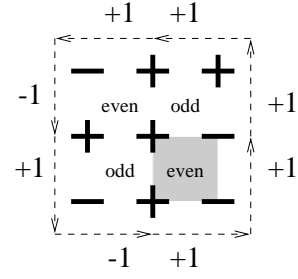

FIG. 8. An example of the sum in equation 5. The charge enclosed is $(1/4) \sum = +1$.

As shown in figure 9, if our single spin-flip dynamics chooses the upper left-hand site of a positive stripe, then the stripe will move by $(-1/2, -1/2)$ plus one-half the sum of $\vec{E}$ on the two edges radiating outward from that site. When the similar statement is made for the other corners, the constants $(\pm 1/2, \pm 1/2)$ cancel, and the average movement of a stripe of charge $q$ is

$$\langle \Delta \vec{x} \rangle = \frac{1}{16} \sum \vec{E}$$

where the sum is over the eight edges radiating out from the stripe's plaquet, shown dashed in the figure. By the same argument as for the triangular antiferromagnet, we argue that there is a first-order force that decreases as $1/r$. Specifically, equation 5 gives

$$\bar{E} = \frac{4q}{2\pi r}$$

if we assume isotropy, and

$$\langle \Delta \vec{x} \rangle = \frac{qq'}{\pi r}$$

in a local mean-field approximation in which $(1/8) \sum \vec{E} = \bar{E}$.

If we look at what combination of forces can occur around a plaquet, figure 10 shows that $\vec{E}$ maps the no-stripes rule onto the six-vertex ice model, where each site of the dual lattice has two incoming and two outgoing arrows. This can also be thought of as a loop-covering model, in which every vertex is covered by exactly one loop [3]. The entropy per site of this model has been calculated exactly by Lieb [25] and is $\frac{3}{2} \ln \frac{4}{3}$.

In this mapping, positive and negative stripes correspond to positive and negative monopoles, with four outgoing or incoming arrows. A single spin-flip corresponds to reversing the four arrows around a site; this preserves the eight-vertex ice rule that each site have an even number of incoming arrows. If we only allow these moves when no new monopoles will be created, they cause monopoles to migrate. Thus we claim that under this dynamics, monopoles in the six-vertex ice model also interact with a Coulomb force.

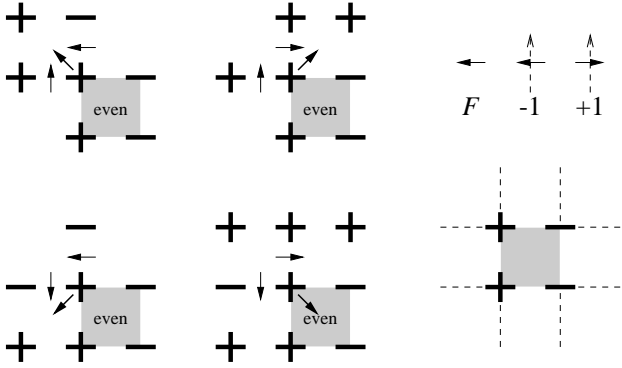

FIG. 9. The average movement of a single stripe is proportional to the average field on the eight dashed edges. Here we show movement resulting from flipping the stripe's upper left site. In the lower-right hand case, the stripe does not move at all, since flipping this site would increase the energy by creating three stripes instead of one.

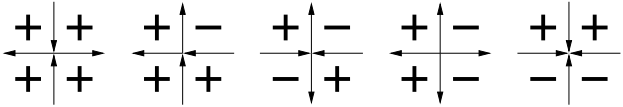

FIG. 10. Using $\vec{E}$ to map the no-stripes rule to the six-vertex ice rule. The plaquets shown are even. Positive and negative stripes become positive and negative monopoles.

Knops points out that the no-stripes rule or the six-vertex ice model can also be mapped to a Villain model [18]. We can then define a vector $w$ on the two sublattices as we did for the triangular antiferromagnet. Let $w$ be 1 and $e^{\pi i/2}$ on odd and even sites respectively. Then $Q$ is the winding number of $s_i w_i$ around the origin,

$$Q = \frac{1}{2\pi i} \oint \frac{d(sw)}{sw}$$

as shown in figure 11, just as for the Ising antiferromagnet. Finally, this model can also be mapped to a body-centered solid-on-solid model [6], in which defects correspond to screw dislocations of Burgers vector $\pm 4$.

## IV. THE $Q = 3$ SQUARE POTTS ANTIFERROMAGNET

The $q$-state antiferromagnetic Potts model is a generalization of the Ising antiferromagnet,

$$U = \sum_{\text{nn}} \delta(s_i, s_j) \qquad (6)$$

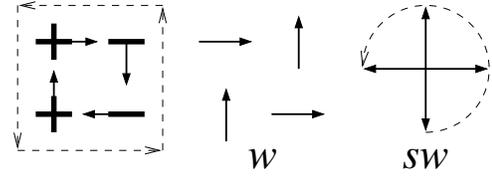

FIG. 11. Defining charge in the no-stripes rule as the winding number of $sw$ around the origin. The upper-left site is even, and the winding of $+1$ corresponds to a positive defect.

where each site has a state $s_i \in \{1, 2, \ldots, q\}$ and $\delta$ is the Kronecker delta function. The states are often thought of as colors, so that the ground state consists of a coloring where no two neighbors have the same color. When $q = 3$, this can be thought of as a discretization of the antiferromagnetic $XY$ model,

$$U = \sum_{\text{nn}} \cos(\theta_i - \theta_j) \qquad (7)$$

where each site has a unit spin pointing in some direction $\theta$. If the $\theta_i$ are restricted to three directions $2\pi/3$ apart, then equations 6 and 7 are equal up to an affine transformation since $\cos(\theta_i - \theta_j)$ depends only on whether $\theta_i$ and $\theta_j$ are the same or different.

We will focus on this three-state model on the square lattice. Baxter [4] and Nightingale and Schick [27] showed that it is critical at $T = 0$, and that this is its only phase transition. Baxter solved the $T = 0$ case as a hard-squares model [5]. For recent Monte Carlo studies using the Wang-Swendsen-Kotecký cluster algorithm [34], see e.g. Ferreira and Sokal [12].

Kolafa [20] pointed out that this model supports vortices as shown in figure 12. These change color but preserve their handedness as they diffuse, and two defects of opposite handedness can annihilate when they meet. There are also chargeless excitations, where sites to either side of the ends of the frustrated bond have the same color. These can be annealed away without interacting with other defects, so they disappear exponentially quickly in a quench.

Kolafa defines the charge within a region as a sum around a counterclockwise perimeter,

$$Q = \frac{1}{6} \sum m(s_{i+1} - s_i) \qquad (8)$$

where $m(k) = 0$, $+1$ or $-1$, and $m(k) \equiv k \pmod 3$. Equivalently, we can define a complex local magnetization $\xi_i = (-1)^p e^{2\pi i s_i/3}$ where the three colors correspond to vectors $2\pi/3$ apart, and $(-1)^p$ gives a sign of $-1$ and $+1$ on odd and even lattice sites respectively [11]. This makes regions with antiferromagnetic order, say with color 1 on one sublattice and colors 2 and 3 on the other, relatively uniform. Then $Q$ is the winding number of $\xi_i$ around the origin,

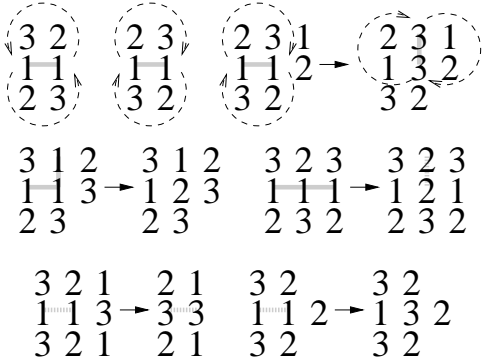

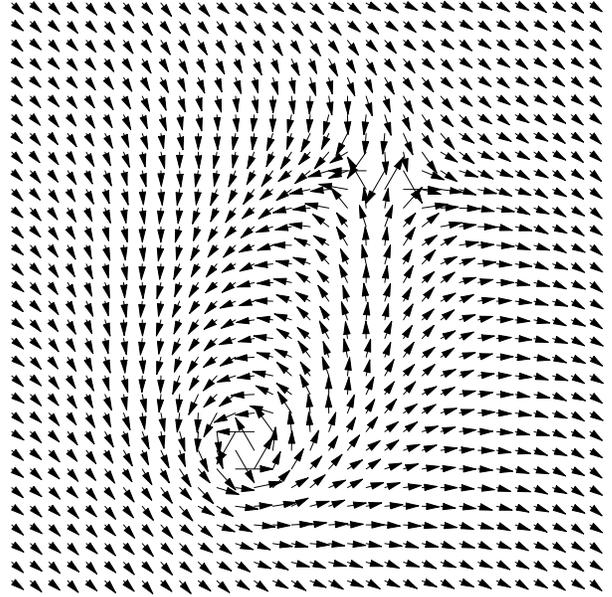

FIG. 12. Counterclockwise and clockwise defects maintain their handedness as they diffuse. Pairs of opposite handedness can annihilate or turn into chargeless excitations, which diffuse until one end is surrounded by only one other color, at which point it can disappear by changing to the third one.

$$Q = \frac{1}{2\pi i} \oint \frac{d\xi}{\xi}$$

In figures 13 and 14, we show the time average of $\xi$ for a configuration with two defects. As with the triangular antiferromagnet, we can clearly see how each vortex produces a field around it which falls off with distance.

Of all these models, this one has the easiest argument for a Coulomb force. As before, we can define a field $\vec{E}$ perpendicular to each bond which will tend to point towards clockwise defects and away from counterclockwise ones. We do this by orienting it so that the higher color in the cyclic ordering $3 > 2 > 1 > 3$ is on its left. As figure 15 shows, after we flip one site of a positive (counterclockwise) defect, the displacement of its midpoint from that site is just one-half the field on the edge extending the frustrated bond. Since both sites are equally likely, the average movement of the defect's midpoint is proportional to $\vec{E}$ averaged over the two bonds extending from the defect,

$$\langle \Delta \vec{x} \rangle = \frac{1}{4} \sum \vec{E}$$

Since we can rewrite Kolafa's formula as

$$Q = \frac{1}{6} \sum_{\vec{b}} \vec{E} \times \vec{b}$$

we again have a first-order $1/r$ force if $\vec{E}$ is sufficiently uncorrelated around a large perimeter. Assuming isotropy gives

$$\bar{E} = \frac{6q}{2\pi r}$$

and

$$\langle \Delta \vec{x} \rangle = 3 \frac{qq'}{2\pi r}$$

FIG. 13. The local magnetization around a pair of vortices in the $q = 3$ Potts antiferromagnet. This picture was obtained by relaxing a $32 \times 32$ lattice until only two defects remained, and then holding them fixed for $10^5$ updates per site while we averaged the magnetization at each site.

in the local mean-field approximation used in the previous two models where $(1/2) \sum \vec{E} = \bar{E}$.

Lenard [24] has pointed out that this field maps the $q = 3$ square Potts antiferromagnet onto the six-vertex ice model. When defects are present, this mapping breaks down as shown in figure 16. The frustrated bond has three outgoing or incoming arrows at each end. Thus defects act like monopoles as in the previous section, but this time composed of a bound pair of vertices with an unoriented bond between them. Nijs et al. [28] defined a 27-vertex model that includes bonds and vertices of this type and studied its scaling properties.

As for the triangular antiferromagnet, the Potts antiferromagnet can also be mapped to a solid-on-solid model where each site is given a height $h_i$ such that $h_i \equiv s_i$ (mod 3) and $h_i$ differs by $\pm 1$ between neighbors. Defects then correspond to a bound pair of screw dislocations with a total Burgers vector of $\pm 6$.

Finally, we note that Bakaev and Kabanovich [2] have discussed the motion of a different kind of defect in the $q = 3$ Potts antiferromagnet, a hole with an undefined color.

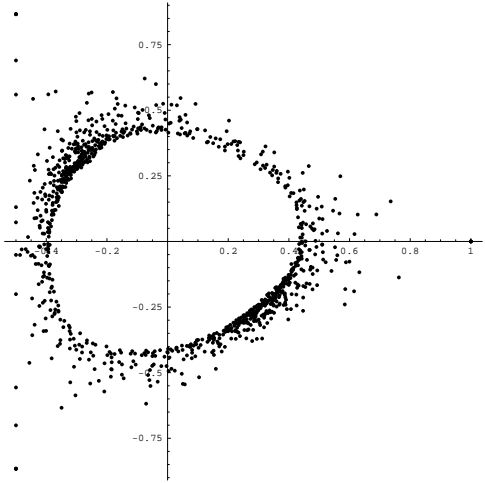

FIG. 14. A scatterplot of the time-averaged local magnetization in figure 13.

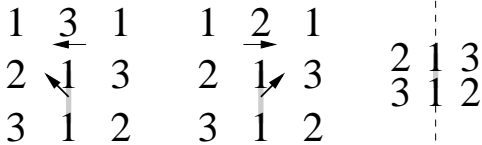

FIG. 15. If we define a field $\vec{E}$ on each bond so that the higher color in the cyclic ordering is on its left, the movement of a positive defect is proportional to $\vec{E}$ averaged over the two dashed edges.

## V. RELAXATION AND DIRECT MEASUREMENTS OF THE FORCE

Yurke et al. [37] discuss the relaxation dynamics of the $XY$ model, in which vortices of opposite type are attracted with a Coulomb force. Assuming a viscous dynamics $\vec{v} = \Gamma \vec{F}$ where the mobility $\Gamma$ is inversely proportional to $\ln r$, they use simple scaling arguments to show that the defect density $\rho$ obeys

$$\frac{1}{\rho^2}\frac{d\rho}{dt} = \frac{C}{\ln(\rho/\rho_c)} \quad (9)$$

where $\rho_c$ is a core density and $C$ is a constant. Note that the left-hand side of equation 9 is constant for the mean-field behavior $\rho \propto t^{-1}$. The leading behavior of $\rho$ as a function of time is then

$$\rho \propto \frac{\ln t}{t} + \mathcal{O}\left(\frac{\ln\ln t}{t}\right)$$

modifying the asymptotic behavior $\rho \sim t^{-1}$ with a logarithmic correction. They confirm the existence of this correction through numerical experiments.

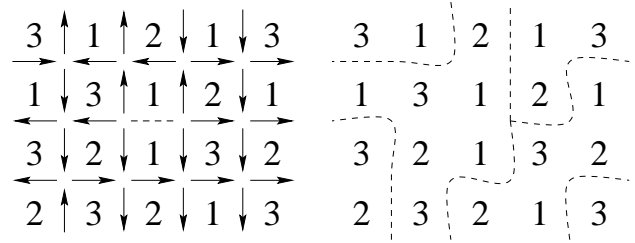

FIG. 16. Mapping the $q = 3$ Potts antiferromagnet onto the six-vertex ice model and the loop-covering model. In the former, the defect corresponds to a monopole, an undefined edge with six arrows emerging from it.

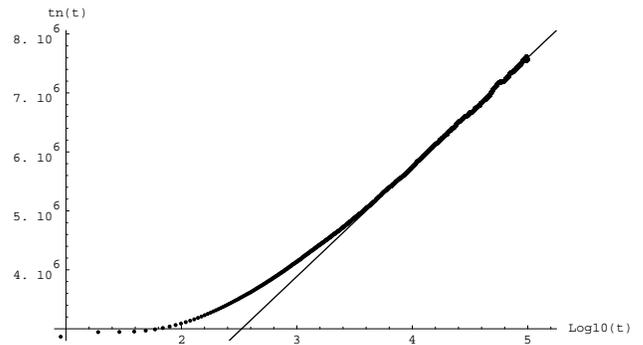

FIG. 17. A plot of $tn(t)$ vs. $\log_{10} t$ for the triangular antiferromagnet. The straight line suggests that $n(t) \propto (\ln t)/t$ for large $t$, and extends about a decade and a half. The data was taken by averaging 100 trials of $10^5$ updates per site on a $4096 \times 4096$ lattice, at which time $\sim 75$ defects remained.

We performed similar experiments for these three models. In each case, we quenched the system from $T = \infty$ (a random initial state) to $T = 0$, and measured the number of defects $n(t)$ as a function of time. In figure 17 we graph $tn(t)$ against $\log t$ for the triangular antiferromagnet, averaged over 100 runs of $10^5$ updates per site each on a $4096 \times 4096$ lattice. While logarithmic corrections are difficult to establish numerically, $n(t)$ seems to behave as $(\ln t)/t$ over one and a half decades in $t$. In addition, in figure 18 we use the same data to graph $\rho^2 (d\rho/dt)^{-1}$ and fit it to $C^{-1} \ln(\rho/\rho_c)$ as per equation 9. We obtain a good fit over a decade and half in $\rho$, with a core density $\rho_c = 0.02$. Results from the other two models are similar.

However, this turns out not to be a good test for a Coulomb force. We have performed a continuous-time free diffusion experiment, in which we use a random state of the triangular antiferromagnet to set up the initial distribution of defects. We then diffuse the defects with random walks, moving them up, down, left and right with equal probabilities, and annihilate pairs of opposite charge when they meet.

Since the total charge in an region of size $l$ is a sum of $\mathcal{O}(l)$ surface terms, rather than $\mathcal{O}(l^2)$ random variables, the total charge fluctuates as $\mathcal{O}(l^{1/2})$ rather than $\mathcal{O}(l)$. Thus defects of opposite charge are well mixed with each other in the initial condition, giving a decay close to the mean-field behavior $\rho(t) \propto t^{-1}$. With uncorrelated initial conditions, defects of like type clump into domains, giving a $t^{-1/2}$ decay [33,9].

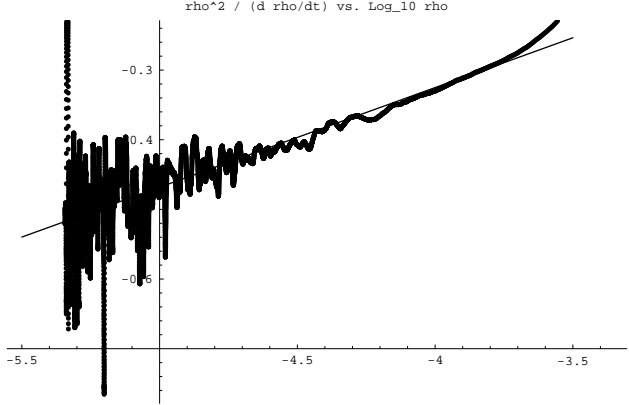

FIG. 18. A plot of $\rho^2/(d\rho/dt)$ vs. $\log_{10}\rho$ for the triangular antiferromagnet, using the same data as in figure 17. The derivative at each time $t$ was defined by a linear fit to 101 data points centered around $t$. The fit is to the form in equation 9. If the mean-field behavior $\rho \propto t^{-1}$ held, the graph would be a horizontal line.

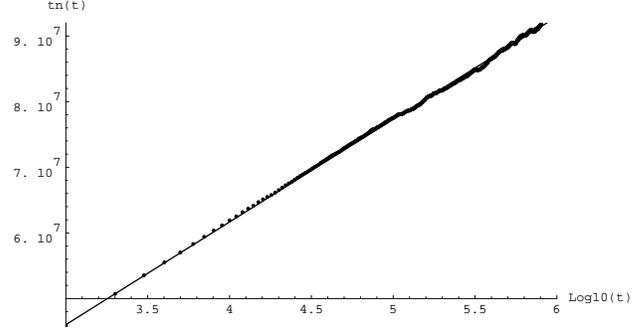

FIG. 19. A plot of $tn(t)$ vs. $\log_{10} t$ in a continuous-time free diffusion experiment, where the triangular antiferromagnet was used to set up the initial distribution of defects. The straight line suggests that $n(t) \propto (\ln t)/t$, and extends three decades. The data was taken by averaging 100 trials of $8 \cdot 10^5$ updates per site each on a $4096 \times 4096$ lattice until $\sim 100$ defects remained.

In figure 19 we graph $tn(t)$ for a $4096 \times 4096$ lattice over a continuous time of $8 \cdot 10^5$ updates per site. The data appears to have the same asymptotic behavior of $(\ln t)/t$, and fits of $\rho^2/(d\rho/dt)$ to $\ln \rho$ are at least as convincing as for our three models. In other words, the asymptotic behavior of $\rho$ is more a consequence of the correlations in the defects' initial positions than of the forces between them. This is presumably because a $1/r$ force causes the length scale of the system to grow asymptotically as $t^{1/2}$, no faster than diffusion would anyway. At short times, however, $\rho$ decreases faster in our three models than in the free diffusion experiment, indicating that attractive forces play a role early on when defects are relatively close together.

Even if the behavior of $\rho$ at long times doesn't confirm the existence of a Coulomb force, we can argue that it does show that the force between defects does not fall off more slowly than $1/r$. Generalizing the argument of [37], if the force between two defects goes as $r^{-\alpha}$ for some $\alpha < 1$ and the mobility is roughly constant, the typical velocity is

$$\frac{d\xi}{dt} \propto \xi^{-\alpha}$$

where $\xi$ is the length scale of the system. Then $\xi \propto t^{1/(1+\alpha)}$, and the defect density is

$$\rho \propto \xi^{-2} \propto t^{-2/(1+\alpha)}$$

If $\alpha < 1$, then, $\rho$ would decay faster than $t^{-1}$. Since we do not observe this, we claim that the field lines spread out, and are not concentrated between defects. (In another paper, in progress, we argue that field lines do bunch into tubes when a ferromagnetic next-nearest-neighbor interaction is added.)

In the case of the $q = 3$ square Potts antiferromagnet, we have also measured the force between defects directly. In an effort to sample the set of configurations with two defects a particular distance away from each other, we set up an initial condition with two vertical defects of opposite type as in figure 20, and then let the lattice evolve while keeping these defects fixed. We did this on lattice sizes of 32, 64, 128 and 256, averaging over $10^7$ updates per site in each case after an initial equilibration of $10^3$ updates per site. Letting the interdefect distance range up to $1/3$ the lattice size, we perform a least-squares fit to the form

$$\langle \Delta \vec{x} \rangle = \frac{A}{r + r_0}$$

where $r_0$ is a core radius. This fit seems to work fairly well.

Our values for $A$ and $r_0$ for various lattice sizes $L$ are

| $L$ | $A$ | $r_0$ |
|---|---|---|
| 32 | 1.142 | 2.573 |
| 64 | 0.978 | 1.956 |
| 128 | 0.906 | 1.652 |
| 256 | 0.869 | 1.494 |

These show considerable finite-size effects. Since the model is critical at $T = 0$, we assume that $A$ converges to its exact value as a power law in the lattice size. A least-squares fit to the form $A(L) = A + CL^\alpha$ gives an exponent of $\alpha = -1.137$ and an extrapolated value of $A = 0.843$. This is larger than the local mean-field approximation $3/2\pi \approx 0.477$ given in the previous section. We claim below that the exact value is $A = 3/4$, in which case our agreement is reasonable but not particularly good. However, this extrapolation is based on only four different lattice sizes, so it should be easy to improve it with further numerical work.

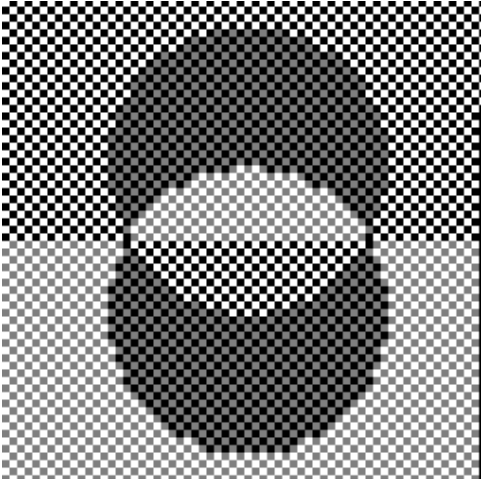

FIG. 20. An initial condition with two vertical defects an even distance away from each other. The six checkerboard phases cycle around the defects, meeting at angles of $\pi/3$.

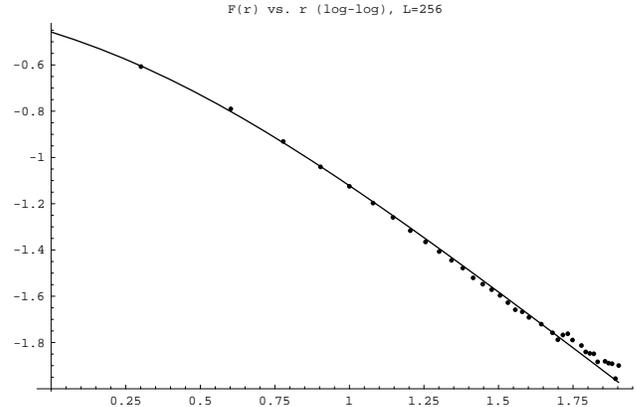

FIG. 21. The force between two defects as a function of their distance $r$, This data was obtained on a lattice of size 256 by starting with the initial condition shown in figure 20 and averaging the force over $10^7$ updates per site after an initial equilibration of $10^3$ updates per site. During these updates, the defects are kept fixed, and $\langle\Delta\vec{x}\rangle$ is defined according to which way they would move if we allowed them to. The form $A/(r + r_0)$ fits the data within the expected error over most of the range.

We suspect that $A$ is larger on smaller lattices because field lines have less room to spread out in. We would like to analyze this in terms of image charges, but it is somewhat unclear how to do this. In fact, even when the interdefect distance is half the lattice size, our prediction of $A/(r + r_0)$ for the force is only 25% off, even though by symmetry the image charges should cancel and make the force zero. The reason for this is that single spin-flip dynamics is not quite ergodic at zero temperature when the defects are held fixed; the boundaries between checkerboard domains in figure 20 can move and join with droplet boundaries, but not cross. Therefore, the topology of the field lines stays the same, with five connecting the defects directly and only one going around the lattice the other way. In a sense, this may be a happy accident, allowing us to observe the force at larger distances than we would be able to on finite lattices if image charges were fully felt. It also means that the measured value of $A$ may depend on the topology of the initial condition; however, experiments on lattices of size 64 and 128 where all six field lines connect the two defects directly give values of $A$ and $r_0$ differing from those above by less than 0.4%.

We also tried to use the Wang-Swendsen-Kotecký cluster algorithm [34] to sample the set of configurations with a pair of defects at particular places. Unfortunately, while cluster-flipping moves leave defects in the same place, they sometimes replace two defects of opposite charge with two chargeless ones. Finding an algorithm more efficient than single spin-flip dynamics to sample this set is still an open question.

## VI. ENTROPIC COULOMB FORCES

In the $XY$ model and other two-dimensional models with continuous degrees of freedom, the energy $U$ of a free defect is typically proportional to $\ln r$ where $r$ is the interdefect distance [21]. Then the energy gradient gives an attractive Coulomb force between defects of opposite type,

$$F = -\frac{\partial U}{\partial r} \propto -\frac{1}{r}$$

In these discrete models, on the other hand, the energy of a pair of defects is 2 regardless of how far apart they are. Therefore, the force must be driven by a gradient in the entropy, or equivalently, by a gradient in the free energy at a nonzero effective temperature.

To show how an entropy gradient can drive a first-order force, suppose that we group the set of spin configurations with two defects into a set of macrostates $\Sigma(r)$, one for each interdefect distance $r$. As the defects move towards and away from each other, we can describe the system as a biased random walk

$$\cdots \rightleftarrows \Sigma(r-1) \rightleftarrows \Sigma(r) \rightleftarrows \Sigma(r+1) \rightleftarrows \cdots$$

If we assume that this walk is a Markov process which has not yet had time to hit the absorbing state $\Sigma(0)$ where the defects annihilate, and if during this transient all microstates with two defects are equally likely, then the ratio of transition probabilities between neighboring macrostates must be the ratio of the number of microstates in them,

$$\frac{P(r \to r+1)}{P(r+1 \to r)} = \frac{\Omega(r+1)}{\Omega(r)}$$

where $\Omega(r)$ is the number of microstates with defects $r$ apart. The average motion is then

$$\langle \Delta r \rangle = P(r \to r+1) - P(r+1 \to r)$$
$$= 2\Gamma \frac{\Omega(r+1) - \Omega(r)}{\Omega(r+1) + \Omega(r)} \approx \Gamma \frac{1}{\Omega} \frac{\partial \Omega(r)}{\partial r} = \Gamma \vec{F}$$

where

$$\Gamma = (1/2) \left( P(r \to r+1) + P(r+1 \to r) \right) \qquad (10)$$

can be thought of as a mobility, and

$$\vec{F} = \frac{\partial S(r)}{\partial r} \qquad (11)$$

is the entropic force where $S = \ln \Omega$.

To get a force proportional to $1/r$, we need to show that the presence of a defect decreases the entropy by an amount proportional to $\ln r$. For the dimer model on the square lattice, Fisher and Stephenson [13] used Pfaffians and Toeplitz determinants to show that the number of configurations with two 'holes' a distance $r$ apart is reduced by $Ar^{-1/2}$, so $\Delta S = -(1/2) \ln r$. Ioffe and Larkin [17] pointed out that this leads to a Coulomb force. In addition, Zeng et al. [39] performed numerical calculations of the cost of dislocation pairs in a randomly pinned fully-packed loop model, and found a cost proportional to $\ln r$. We will give several arguments for a similar term in the models discussed here.

One simple counting argument goes as follows. Suppose we are trying to extend a spin configuration outward from a square region of the lattice by adding an additional layer of $l$ sites around its perimeter. For the Potts model, for instance, successive sites have colors $s_i$ differing by $\pm 1$, and the sum in equation 8 is the total displacement of a walk of length $l$. The number of such walks is

$$\binom{l}{l/2 - 6\,Q}$$

where $Q$ is the charge inside the perimeter. The larger $Q$ is, the greater the constraint on the walks, and the lower the entropy will be. If we ignore the interaction between successive layers, which of course we can't, then the entropy of the set of states surrounding a charge $Q$ (we take Boltzmann's constant to be 1) is

$$S_Q \approx \ln \prod_{l=0}^{r} \binom{l}{l/2 - 6\,Q}$$
$$= \sum_{l=0}^{r} \ln \binom{l}{l/2 - 6\,Q}$$
$$= S_0 - \sum_{l=0}^{r} \left( \ln \binom{l}{l/2} - \ln \binom{l}{l/2 - 6\,Q} \right)$$
$$\approx S_0 - (6\,Q)^2 \sum_{l=0}^{r} \frac{1}{l}$$
$$\approx S_0 - A\,Q^2 \ln r$$

for some constant $A$. We cut off our sum at some maximum perimeter proportional to the interdefect distance $r$, outside which the charge $Q$ is cancelled by defects of the opposite type.

While the assumption of independent layers is hugely wrong (for instance, it gives a ground state entropy per site of $s_0 = S_0/N = 2$) we find that the presence of a charge $Q$ reduces the entropy by an amount $\Delta S$ proportional to $-Q^2 \ln r$. We can think of this as a contribution to an effective free energy,

$$G = -\Delta S = A\,Q^2 \ln r$$

This matches nicely with the energy of a dislocation with Burgers vector $B$,

$$U = CB^2 \ln r \qquad (12)$$

where $C$ is a constant dependent on the lattice spacing, the rigidity modulus, and Poisson's ratio [14].

As another approach, Kotecký [22] points out that the entropy of the Potts model can be related to the energy of a ferromagnetic Ising model at a nonzero effective temperature, since the entropy of one sublattice is higher if sites on the other sublattice are the same. An odd site (say) has two choices of color if its four neighbors have the same color, or equivalently if the field $\vec{E}$ points clockwise or counterclockwise around all four bonds, and only one choice otherwise. Recall that the field $\vec{E}$ determines whether the color changes by $+1$ or $-1$ between neighbors. If we define the probability of the color increasing in the $x$ and $y$ directions as $p_x$ and $p_y$ respectively, we have

$$p_x = (1 - E_y)/2$$
$$p_y = (1 + E_x)/2$$

where $E_x$ and $E_y$ are the components of the average field. If $\vec{E}$ is slowly varying, and if the colors of the four neighbors are independent, which they aren't, the probability of all four having the same color is

$$2p_x(1-p_x)p_y(1-p_y) = \frac{1}{8}(1-E_x^2)(1-E_y)^2 \approx \frac{1}{8}(1-E^2)$$

for small fields $E \ll 1$. The average entropy per site is

$$s = \frac{\ln 2}{8}(1-E^2) = s_0 - \frac{\ln 2}{8}E^2$$

where $s_0$ is a (badly underestimated) ground state entropy of $(1/8)\ln 2$ per site. The effective free energy is then increased by

$$G = \frac{\ln 2}{8}\int E^2\, dx\, dy \tag{13}$$

giving an energy density proportional to the square of the field, just as for electromagnetism.

To make this more precise, we can use the height representation for these models [28], in which there is an effective free energy

$$G = \frac{1}{2}K\int |\nabla h|^2\, dx\, dy$$

Note that $\nabla h$ is simply $\vec{E}$ rotated counterclockwise by $\pi/2$, so $|\nabla h|^2 = E^2$ and this has the same form as equation 13. A screw dislocation with Burgers vector $B$ has a field around it

$$|\nabla h| = |\vec{E}| = \frac{B}{2\pi r}$$

so the free energy of a defect integrated from a short-distance cutoff $r_0$ (roughly the lattice spacing) to an interdefect distance $r$ is

$$G = \frac{KB^2}{4\pi}(\ln r - \ln r_0)$$

which again matches the form in equation 12. The force between two screw dislocations with Burgers vector $B$ and $B'$ a distance $r$ apart is then

$$F = -\frac{\partial G}{\partial r} = K\frac{BB'}{\pi r} \tag{14}$$

Combining this with equation 11 gives

$$\langle \Delta r \rangle = \frac{\Gamma K B^2}{\pi}\frac{qq'}{r} \tag{15}$$

where $B$ is the magnitude of the Burgers vector for a defect of charge $\pm 1$.

For the $q = 3$ Potts antiferromagnet, we can calculate the coefficient $A = \Gamma KB^2/\pi$ of equation 15 exactly. Park and Widom [30], using the exact solution of the six-vertex ice model, showed that the free energy of an interface of width $L$ across a height difference of $\Delta h = \pm 2$ is $2\pi/6L$, and Burton and Henley [10] point out that setting this equal to $(1/2)K(\Delta h/L)^2 L$ gives $K = \pi/6$.

To calculate $\Gamma$, we need to take into account the fact that a diffusing defect alternates between horizontal and vertical bonds. If we assume for simplicity that $\vec{E}$ is in the $x$-direction, a vertical defect which flips to a neighboring horizontal bond has an equal probability of flipping back to its original position or moving to the vertical bond one site away. If $P_\to$ and $P_\leftarrow$ are the probabilities that it will move to the horizontal bond to its right and left respectively, after two time-steps we have $P(r \to r+1) = P_\to/2$ and $P(r+1 \to r) = P_\leftarrow/2$. Since $P_\to + P_\leftarrow = 1$, plugging these into equation 10 and dividing by 2 since $\Gamma$ is inversely proportional to time gives $\Gamma = 1/8$.

Combining these with $B = 6$ gives $A = 3/4$. Our experimental value of 0.841 differs from this by 12%, but this extrapolation was based on only four lattice sizes. We hope to measure $A$ more accurately in the future.

## VII. CONCLUSION

We have examined topological defects or vortices in the triangular Ising antiferromagnet, a model with two- and four-point interactions on the square lattice equivalent to the six-vertex ice model under a dynamics where elementary moves reverse the four arrows around a plaquet, and the three-state Potts antiferromagnet on the square lattice. In each case, positive and negative defects appear to be attracted by a Coulomb force, and we have confirmed this through numerical experiments. For the Potts model, we have obtained reasonable agreement between the force coefficient and its predicted value.

Both the triangular Ising antiferromagnet [23,26] and the $q = 3$ Potts antiferromagnet [29] are known to have Kosterlitz-Thouless-like phase transitions [21] in the same universality class as the six-state clock model [19,36] when a ferromagnetic next-nearest-neighbor interaction is added. In another paper, we will look at how this interaction, and the screening effect of particle-antiparticle pairs at nonzero temperatures, affects the forces between defects. We will also look at vortex loops in various three-dimensional models.

**Acknowledgements.** C.M. is grateful to Mark Newman, Chris Henley, Chen Zeng, Roman Kotecký, Michael Lachmann, and Andrew Pargellis for helpful discussions; Molly Rose and Spootie the Cat for inspiration; and Margaret Alexander for unfailing bibliographic support. N.M. thanks the National Science Foundation Research Experience for Undergraduates program. We also wish to express heartfelt sympathy for users of `xfig` everywhere.